# Tempo-Invariant Processing of Rhythm with Convolutional Neural Networks

Anders Elowsson

## Abstract

Rhythm patterns can be performed with a wide variation of tempi. This presents a challenge for many music information retrieval (MIR) systems; ideally, perceptually similar rhythms should be represented and processed similarly, regardless of the specific tempo at which they were performed. Several recent systems for tempo estimation, beat tracking, and downbeat tracking have therefore sought to process rhythm in a tempo-invariant way, often by sampling input vectors according to a precomputed pulse level. This paper describes how a log-frequency representation of rhythm-related activations instead can promote tempo invariance when processed with convolutional neural networks. The strategy incorporates invariance at a fundamental level and can be useful for most tasks related to rhythm processing. Different methods are described, relying on magnitude, phase relationships of different rhythm channels, as well as raw phase information. Several variations are explored to provide direction for future implementations.

## 1. Introduction

### 1.1. Invariant processing of rhythm in the time domain

*Tempo invariance*
Musical rhythm and meter establish a perceptual framework within which the listener can interpret musical sounds and structure. The musical meaning of a note onset is determined by its timing relative to other notes, and, ultimately, its position within the metrical structure. For example, a note onset 0.125 seconds (s) after the start of a new measure will have a different meaning for a song with a tempo of 120 beats per minute (BPM) (corresponding to a $16^{th}$ note after the start of the measure) and 160 BPM (dotted $8^{th}$ note after measure start). It is therefore not surprising that the music notation system is based on note values relative to the tempo of the song, instead of absolute time. Tempo is perceived logarithmically to a large extent. For instance, an increase from 40 BPM to 80 BPM is perceived as a doubling, just like an increase from 60 BPM to 120 BPM. This can be used in rhythm processing systems. For example, random fluctuations of the next beat positions can be allowed to vary according to Gaussian distribution over the logarithm of the inter-beat interval of the previous beat (Korzeniowski et al., 2014); and tentative tempi candidates can be associated by their BPM-ratio (Elowsson and Friberg, 2015).

*Beat invariance*
By using the metrical framework during MIR processing, a system may more efficiently interpret musical sounds. The system proposed by Krebs et al. (2016) computes input features for downbeat tracking aligned with previously computed beat positions. A similar strategy is used by Papadopoulos and Peeters (2011). Durand et al. (2015; 2016) also computed pulse-synchronous features for downbeat tracking, opting instead to use subdivisions of the beat to ensure a high recall of potential downbeat positions. They report a recall of "above 95 %" using a 100-millisecond (ms) window. Beat-invariant chord estimation systems have also been proposed (Mauch et al., 2009), as well as beat-aligned chroma vectors for artist recognition (Bertin-Mahieux, Weiss, & Ellis, 2010).

*Cepstroid invariance*

Beat-invariant systems use both tempo and phase invariance at the beat level. However, it is harder to estimate beat positions than to estimate the periodicity, which results in more examples with incorrect initial conditions for the subsequent processing. Therefore, the bar-pointer model proposed by Whiteley, Cemgil, and Godsill (2006) and derivative implementation (Krebs, Böck, & Widmer, 2015) tries to permute the state space by jointly estimating tempo and beat positions. State-space models have also used visual information in combination with audio aspects (Ohkita et al., 2017). Another strategy is to compute the "cepstroid," corresponding to the most salient level of repetition in the music. It is computed by applying the discrete cosine transform to an interonset histogram (Elowsson and Friberg, 2013; Elowsson and Friberg, 2015). Invariance can then be achieved by subsampling the input to the network with a hop size derived from a subdivision of the computed cepstroid (Elowsson, 2016). As it is easier to identify the most salient level of repetition than to identify beats, this can introduce fewer errors in the initial step.

*Definition of rhythm invariances*

There are various definitions of invariances in rhythm processing. For example, systems that are invariant with regard to the phase of the rhythm have often been described as "shift-invariant". However, many representations described in this article facilitate shift invariance in the dimension across which, for example, convolution is applied. Therefore, the relevant dimension or music concept will be referred to instead, as outlined in Table 1.

| **Invariance** | **Short description** |
|---|---|
| Tempo-invariant | Invariant with regard to tempo. |
| Phase-invariant | Invariant with regard to the phase of the rhythm at a specific point in time. |
| Rhythm-invariant | Tempo-invariant and phase-invariant. |
| Beat-invariant | Invariant representations computed by subsampling data according precomputed beats. |
| Cepstroid-invariant | Invariant representations computed by subsampling data with a hop-size derived by a salient periodicity. |
| Time-invariant | Invariant across time, not directly related to the phase of the rhythm. |
| Pitch-invariant | Shift-invariant properties of pitch across frequency. |

**Table 1.** Invariances that are useful for processing music rhythm.

## 1.2. Rhythm processing in the frequency domain

*Spectro-temporal representations of rhythm*

The periodicity of rhythmical events can be computed from the discrete Fourier transform (DFT) of onset activations. This produces a linear-frequency spectrum of periodicities. A comprehensive review of rhythm processing strategies in the frequency domain is provided by Quinton (2017). That work is primarily related to tracking metrical modulations. Other researchers have computed the linear-frequency rhythm spectrum from the DFT for tempo and beat tracking (Peeters, 2005a), rhythm classification (Peeters, 2005b), and metrical analysis (Robine, Hanna, & Lagrange, 2009). The autocorrelation function can also be applied to an onset activation signal with the same purpose (Scheirer, 1997; Gouyon & Herrera, 2003; Dixon, Pampalk, & Widmer, 2003). A recent study estimated tempo, dance-style and meter by applying oscillators to rhythm activations, extracting the response at different frequencies (Gkiokas, Katsouros, & Carayannis, 2016). On a similar theme, Schreiber and Müller (2017) use 50 features for tempo octave correction capturing the energy at 10 beat periodicities (log-spaced) × 5 spectral bands.

In the two studies by Peeters (2005a; 2005b), and in later work by Quinton (2017), the autocorrelation is used in conjunction with the DFT to compensate for shortcomings in both functions related to peaks not corresponding to salient (perceptually relevant) metrical levels. Authors have referred to spectro-temporal representations of rhythm as "metergrams", "rhythmograms" and "tempograms".

Many of the systems described above use phase-invariant representations of rhythm; the phase component from the frequency transform is simply discarded. Phase-invariant frequency representations can be particularly useful for pattern matching (Klapuri, 2010). Most of the above-mentioned methods are, however, focused on extracting rhythmical information in the linear-frequency domain. Such a representation does not allow for tempo-invariant processing across frequency through the weight-sharing mechanisms and convolutions incorporated in standard convolutional neural network (CNN) architectures.

The scale transform (Cohen, 1993) has been applied to various signals describing rhythm to extract tempo-invariant features. The first work was done by Holzapfel and Stylianou (2009; 2011), who applied the scale transform to classify dance music. Marchand and Peeters (2014; 2016) subsequently applied the scale transform across several frequency bands to extract modulation scale spectrum (MSS) features.

## 1.3. Rhythm processing with CNNs

Convolutional neural networks for processing rhythm have so far mainly been employed for their time-invariant properties (i.e., events at different times can be filtered by identical kernels), and in some cases for their pitch-invariant properties (i.e., similar events at different frequencies can be filtered by identical kernels). Networks have been proposed for onset detection (Schluter & Böck, 2014) and downbeat tracking (Durand et al., 2016). A shallow CNN with filters of different widths across time has been proposed for event detection (Phan et al., 2016), later extended by Pons and Serra for processing rhythm (2017). Architectures with CNNs have also been employed in multimodal contexts; for a dancing robot (Gkiokas & Katsouros, 2017), or to incorporate visual information (Ohkita et al., 2017). Time-invariant neural networks can of course also be implemented as feedforward neural networks, shifted and applied to each frame (Elowsson, 2016).

## 1.4. Processing rhythm in the log-frequency domain

The methodology proposed in this paper is an attempt to extend previously mentioned work with a generalized tempo-invariant processing architecture facilitating weight sharing. Many of the beat-invariant and cepstroid invariant systems described in Section 1.1 divide the learning into several steps. The invariance is encoded in the pre-processing for the subsequent step. This means that any errors introduced in the pre-processing step will be propagated to subsequent steps. Furthermore, the systems cannot adjust parameters of the earlier steps via gradient descent to improve the performance of later steps. There are situations related to rhythm processing where this may hamper performance. For example, certain low-level features related to onset characteristics may be important in later steps; it is often hard to anticipate how various features are useful. Is there any way to encode rhythm or tempo invariance in the weight sharing of a CNN? If so, the two steps could potentially be combined into an integrated learning system of musical rhythm.

Various metrical levels are generally associated at a logarithmic scale – the tatum level can be described in reference to the beat, commonly being 2, 3, or 4 times faster, and there are generally 2, 3, or 4 beats per measure, etc. Rhythmic invariance, therefore, requires that properties related to timing (such as two onsets) are processed or filtered at a logarithmic time scale. Furthermore, the repetitive structure of rhythm in music means that the frequency and phase of rhythmical elements can represent the rhythm in a perceptually valid, and effective, way. These conditions taken together suggest that a log-frequency representation of temporal signals that capture rhythmically relevant music information could be used in one or several layers of the network. Such a log-spectral rhythm representation could either be computed as a pre-processing step and then used for the input to the network (computed for example from spectral bands), or it could be computed within gradient descent at a hidden layer of the network. Convolution across frequency, which can be implemented with existing CNN architectures, will in either way become tempo-invariant. Max-pooling can be applied to the convolved representations, e.g., to compensate for the shift in frequency imposed on the log-frequency representation for music with different tempi.

A log-frequency rhythmogram was computed by Elowsson (2016). In that study, the rhythmogram was filtered to only retain energies within an octave of a detected tempo, so that when the rhythmogram was inverted back to the time domain, beats appeared as peaks of sinusoidal curves across time. A rhythmogram has also been converted from linear frequency to log-frequency by Quinton (2017); for tracking metrical modulations with non-negative matrix factorization (NMF). With this strategy, however, the rhythmogram peaks will not have the same shape across frequency due to the properties of the linear-frequency transform. This was partially (and implicitly) compensated for by median filtering across frequency. Jensen et al. (2009) computed the autocorrelation function of a high-pass filtered onset signal and smoothed the response with log-spaced filters; the output was used for rhythmic pattern matching.

## 1.5. Overview and organization

This article presents how rhythm-invariant deep learning can be used by processing temporal signals of rhythm accents in the log-frequency domain. It is intended to provide a vision and direction for potential future implementations. Section 2 gives an overview of the processing steps, using only magnitude information in the processing. This methodology can be used, for instance, for tempo estimation and analysis related to metrical aspects. Omitting phase however limits the functionality somewhat, especially in how to connect the log-frequency processing back to the time domain. In Section 3, phase information from the frequency transform is added to the described processes. A way to feed relative phase to the convolutional layers is described, which enables the network to model interrelationships between rhythmical elements (Section 3.2). Section 3.3 describes a method for using the phase when connecting back to the time domain, useful for beat tracking and downbeat tracking. Some techniques for increasing generalization are proposed in Section 3.4. Section 3.5 explores how time domain targets can be used during training, proposing that annotations can undergo a log-frequency transformation identical to that of the input data. As a result, training can more easily be performed in the frequency domain. Section 4 briefly explores how phase could be used in a complex-valued network with complex-valued neurons. This methodology can be useful for all the aforementioned rhythm tasks. The tables in Section 5 summarize relevant rhythm tasks and some additional remarks are offered in Section 6.

The modifications and improvements in the different Sections can be used as a guide for how to iteratively develop a framework for music rhythm intelligence (i.e., Section 2 describes a processing chain useful for a smaller subset of rhythm tracking tasks and Section 3 extends the scope).

## 2. Convolutional neural network model

### 2.1. Building blocks

Outlined in this Section are basic building blocks for tempo-invariant processing using magnitude spectra. The log-spectro-rhythmical representation enables the system to process tracks of different tempo with the same filters, to discover, for example, the tempo, the metrical structure, or similar features. Figure 1 shows two similar rhythm tracks of varying tempo processed by a tempo-invariant CNN. The intention is to illustrate inputs, intermediate representations, filters, and outputs, and how various processing steps are connected. In the example, the constant-Q transform (CQT) (Schörkhuber et al., 2014) was applied to percussive activation signals, and pre-specified filters applied (the filter kernels were not learned through gradient descent in the example).

*Overview*

In step **A**, rhythm activations across time are computed; activations for two examples are shown in pane **B**. Step **C** transforms the activation channels to the log-frequency domain, shown in pane **D** for *one time frame*. Steps **E** and **F** consist of one or more convolutional layers (including activation functions and max-pooling) that process music in a tempo-invariant way. Pane **G** illustrates that similar weights have been computed for the two music examples, as they had the same rhythmical structure. In step **H**, fully connected layers can be applied, if suitable for the task. The subsequent part of this Section (2.1) gives a more detailed overview of each step/pane (A-H) in Figure 1. Potential tasks that can be solved by the method are outlined in Section 2.2.

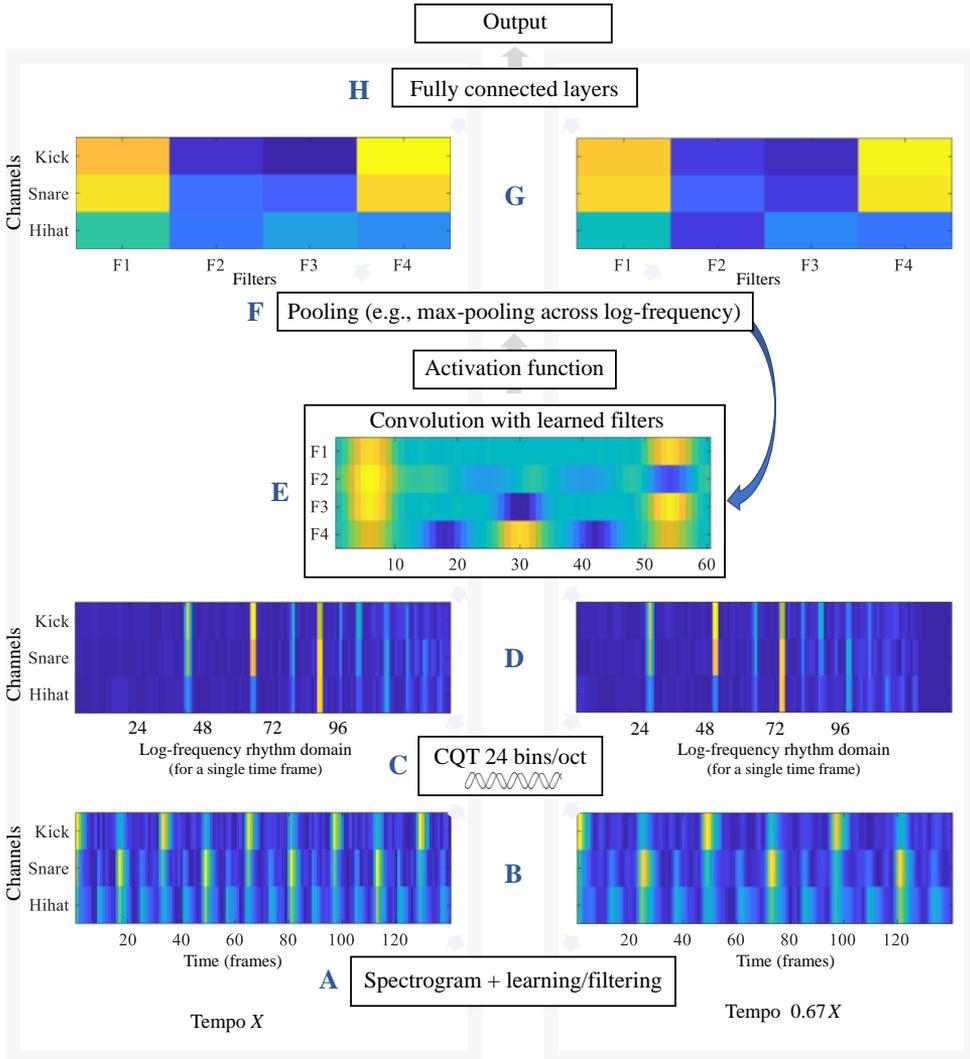

**Figure 1.** Building blocks for a tempo-invariant processing of music with CNNs. The figure shows two drum rhythms of different tempo processed in a tempo-invariant way through steps A-H. The example extracts features representing periodicity patterns in three different drum instruments.

**A**

In the first step (A), aspects of audio relevant to the perception of rhythm are detected in each time/frame. Such aspects can be, for example, chord changes, pitched onsets, or other onsets from various percussive sounds. An activation vector is computed for each aspect across time; these vectors, as well as their associated time-frequency representation, will be referred to as *channels* throughout the rest of the article. How many channels should be used? Most implementations that subsequently compute a frequency representation of rhythm use a single channel consisting of an onset activation vector (see Section 1.2). With more channels it becomes possible to capture different types of periodically related rhythm information in the audio – e.g., different drum parts, or pitched onsets in different octaves. An upper bound is set by factors that affect generalization such as the size of the training set, the variety of periodically recurring aspects in the music style, and computational resources; less than 200 channels should probably be suitable for most situations.

The design of step A will depend on if the whole system is integrated under gradient descent (referred to as *end-to-end learning* in this article) or if intermediate targets are used before step E (referred to as *layered learning*). Signal processing strategies could also be employed. In either case, it is assumed that the first step in A is to extract a time-frequency representation of the audio.

*End-to-end learning through A*

If step C is encoded within gradient descent, the system will learn how to extract the channels through backpropagation, from step F or H to step A. In this case, step A consists of one or several, e.g., convolutional layers, processing the time-frequency representation of the audio. The design of these initial layers is non-trivial, so a few variations are outlined below.

Basic filters could be designed to capture pitched onsets or percussive onsets, for instance. Figure 2 shows a receptive field that extracts pitched onsets from a log-frequency spectrogram – to give an example of the types of receptive fields that could be learned. The receptive field in the example is formed from many copies of a local basis filter, computed by the outer matrix multiplication of a *difference of Gaussians* (across pitch) and the *first-order derivative of a Gaussian* ($g_t$) (across time). Each basis filter is multiplied by a sparse set of weights and inserted at specific relative frequency bins, using learned weights inspired by the $f_0$ estimation step proposed by Elowsson (2018a). Onset detection from the $g_t$ has been proposed by, for instance, Lindeberg and Friberg (2015). Onsets are extracted at frequency bin 0, assuming 60 bins per octave in the spectrogram.

The $g_t$ could be a hard-coded feature for each frequency bin of a filter kernel, to restrict the number of parameters (assuming end-to-end learning, it will be necessary in many scenarios to keep the number of parameters low at this part of the network). During backpropagation, the network would then use all computed gradients across time for a specific frequency bin to determine how the $g_t$ should be tilted, while retaining its shape. A $g_t$ tilted so that it is positive in the second half will activate for increases in the energy of a partial, while the opposite instead suppresses onsets, useful for "inter-partials" as defined by Elowsson (2018a). To further restrict the number of parameters, sparsity should be enforced, i.e., less relevant relative frequencies can be excluded.

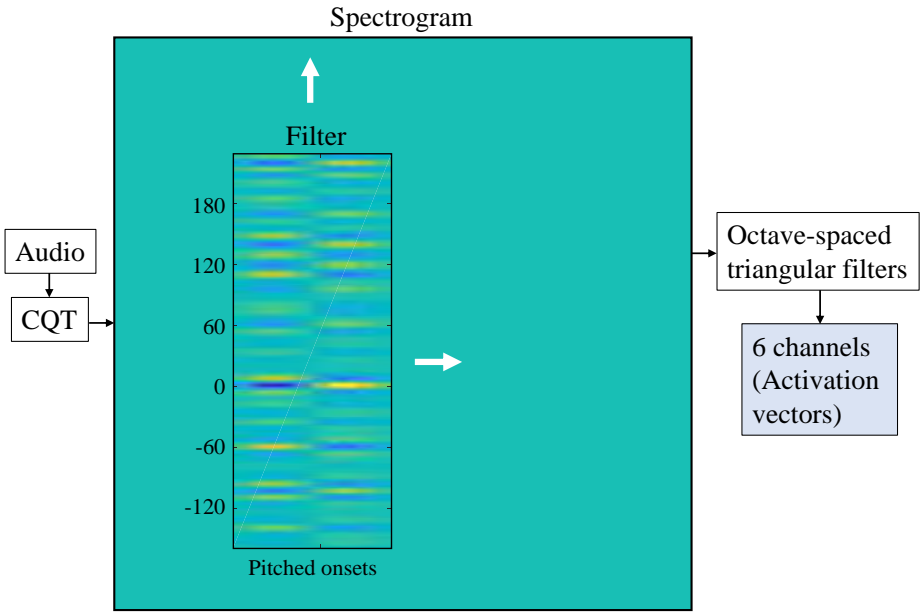

**Figure 2.** A simple overview for how channels for pitched onsets can be computed in step A. The shown filter is convolved across time and frequency of a log-frequency spectrogram, and octave-spaced triangular filters are applied to extract 6 channels of pitched onsets (one channel per octave).

Other receptive fields may be more useful for capturing various percussive onsets in different channels. Smooth receptive fields could extend across the whole spectrum, using one receptive field for each relevant percussive aspect. One or a few smooth filters with a width of around two octaves could convolve across frequency, using a stride of about an octave, so that each filter output corresponds to a separate channel. It is reasonable to assume that hard-coded aspects of the filters (as described for pitched onsets) could be useful for promoting generalization. The various receptive fields could be preceded by layers for harmonic/percussive source separation (HPSS). It has previously been found that the percussive part of HPSS can be very useful for tempo estimation and beat tracking (Gkiokas et al., 2012; Elowsson & Friberg, 2015). Although the subject is important, further suggestions are beyond the scope of this article.

*Layered learning in A*
If step C is not encoded within gradient descent, the system can use intermediate targets to compute relevant channel activations, i.e., layering the learning into separate steps (Elowsson, 2018b). A requisite is therefore that there exist relevant annotations for low- to mid-level concepts that channels should activate for. Fortunately, such annotations are widely available, in the form of annotated onsets, pitched onsets, chord changes, or drum parts, etc. Available pre-trained systems with a high performance can also be applied, e.g., for polyphonic pitch tracking (e.g., Elowsson, 2018). Toolboxes such as Madmom (Böck et al., 2016) also have pre-trained systems of high quality readily available. The benefit of the deep layered learning strategy is that high-quality systems with known functionality can be employed with little work. Furthermore, it may discourage the system from making predictions based on confounding irrelevant factors in the training set. The downside is that rhythmic concepts that are important but hard to envision beforehand may not be captured very well.

*Signal processing strategies in A*
It may also be useful to apply filtering strategies in step A to extract low-level information in channels that do not directly correspond to music concepts. The spectrum can be divided into frequency bands (e.g., *Bark scale critical bands*), or Mel-frequency cepstral components (MFCCs) can be extracted. Another option is to apply generalized receptive fields (Lindeberg & Friberg, 2015), i.e., not learned but pre-defined receptive fields.

**B**

Pane B shows two examples of channels computed in step A. Each example consists of three channels that mostly activate for kick drum, snare drum and hi-hat respectively. These activations can be the result of a filtering strategy in step A focusing on percussive bass, low-mid, and high frequencies, or the result of a layered learning step for drum transcription. They can also have been inferred in a hidden layer within gradient descent if interactions between these parts are relevant to the specific overall task. The left example was created to represent a basic 4/4-meter drum rhythm, with 15 % noise added. The right example was a copy of the left drum rhythm, also containing 15 % noise (different from the first) and upsampled to have 2/3 of the tempo of the first example. Both examples were created to extend about 17 measures.

**C**

Step C is the log-frequency transformation, which is applied to the channels, resulting in a rhythmogram *for each channel*. The transformation is applied, just like the short-term Fourier transform (STFT), to short overlapping segments of each channel, extracted from the longer channel vector across time. The CQT is, however, different from the STFT in some regards. It uses a varying window width, with longer windows for lower frequencies. This gives individual peaks in the frequency spectrum the same shape (width). Note that frequency represents periodicity across time, i.e., we are dealing with "periodicity frequencies." The transform should be restricted to relevant frequencies, where the lowest frequency picks up repetition at or beyond the measure level, while the highest frequencies respond to repetition at $16^{th}$ notes, or slightly higher. This requires window sizes of about 3-5 seconds (picking up at least two measures of music for common slow tempi), similar to the window sizes previously used for beat tracking (Elowsson, 2016). For the phase analysis in Section 3, it is assumed that the transformation will be "centered," i.e., that the phase reference will correspond to the middle of the windowed signal.

For the example in Figure 1, the transformation was computed with a Matlab implementation of the CQT (Schörkhuber et al., 2014). The frequency range was set to 1-1000 Hz with a sampling frequency of 100 Hz, using 24 bins/octave. The CQT was computed for the complete vector and the middle time frame was used in pane D to show the frequency response. Parameters of the CQT should be tuned to handle tempo drifts while still getting defined peaks in lower frequencies. For the end-to-end deep learning system, the signals and their gradients will need to be propagated through the log-frequency transform. This has been done in many implementations for the DFT (Rippel, Snoek, & Adams, 2015). Although there is no clear theoretical restriction as to why this cannot be done for the CQT, no examples were found in a literature search. Such an implementation is, however, beyond the scope of this article. The CQT is slower to compute than the DFT, which can use the speed of the fast Fourier transform (FFT). Fortunately, the FFT can be used in conjunction with a kernel to compute the CQT (Brown & Puckette, 1992).

**D**

The magnitudes of the log-spectral representation of rhythmic repetition are shown, *for a single time frame*, for the two rhythm examples in pane D. Before the log-frequency transform, the music was represented by two dimensions: channels and time (although if a CNN was used prior to step C, channels might have been organized into two dimensions, including "depth"). After the log-frequency transform, the music is instead represented by three dimensions: channels, time, and frequency. In other words, each channel has a spectral representation in each time frame.

The rhythm-invariant properties of the representation are clearly visible in pane D: the right example is very similar to the left example, just shifted 14 bins lower in frequency because of the slower tempo. The lowest frequency corresponds to repetition at half the measure level (one kick drum and one snare drum in a regular beat in 4/4 meter). The next octave corresponds to the beat level. This is where the snare and kick channel respond the strongest, due to the leakage between kick and snare in the input shown in pane B. The hi-hat channel responds strongest at eight notes, two octaves above.

**E**

The log-frequency representations are processed with filters using weights learned through gradient descent. The extent of the filters and their convolutional operation will vary depending on the task.

*Across frequency*
In pane E of Figure 1, four filter kernels were specified by hand. The shown filters extend (only) across frequency. In the example in the figure, the filters facilitate tempo-invariant processing through weight sharing and convolution across frequency. Convolution across frequency, in at least one layer, is a necessity for tempo invariance.

*Across channels*
Filters can also extend across channels. This is beneficial for most tasks in conjunction with convolution across frequency, as it enables the network to compare rhythmic periodicities from different sources in a tempo-invariant way. This comparison can happen in step H instead, but not in a tempo-invariant way.

Filters can also use weight sharing and convolution across channels, whenever channels encompass locality. For example, channels may consist of onsets at different pitch ranges. Convolution across channels can then be used to process relative pitch relationships with the same weights at varying pitch. Another example is if channel activations were computed as spectral fluctuations in different frequency bands. By convolving across channels, rhythmic periodicities at {50, 100, 200} Hertz (Hz) can be compared in the same way as rhythmic periodicities at {100, 200, 400} Hz. To understand why this may be beneficial, listen to a music track that undergoes a pitch shift. The perceived rhythmical accents (downbeats, beats, etc.) will stay the same across several octaves; listeners use relationships between sounds of different frequencies rather than absolute frequencies to establish a rhythmical framework.

There is an important phase component related to the *interaction between various rhythmical elements*, as discussed in Sections 3.1. Furthermore, for all instances where the purpose of the network is to activate at specific events, such as a beat position, the phase will also be a necessary requisite for *associating a specific frequency with a specific point in time*. This use of phase is explored in Section 3.4.

*Across time*

Filters that extend across time could potentially be used to identify tempo drifts, variations in rhythm (for example between different parts of a song), or for smoothing across time (however, note that each time frame may represent periodicities across several seconds in lower frequencies). Convolution, and thereby weight sharing, will also be performed across time because the same processing should be applied to all parts of the music in most cases.

*Across filters*

It may also be beneficial to have filters that extend across the depth of all filters in a previous layer if the convolutional layer is repeated several times. This may be especially beneficial in the last of several such convolutional layers, for concatenation.

F

After non-linear activation functions have been applied, a pooling operator can be used as a final step of each convolutional layer. Only a single convolutional layer is shown in Figure 1 (step D-E-F), but it is likely that more than one layer will provide the best results, with non-linear activation functions in between. In this case, the layer (step D-E-F) can be repeated several times, as indicated by the arrow in the figure.

*Across frequency*

The output of step E will be very similar, but frequency-shifted, for, e.g., identical rhythmic patterns of different tempi. For many tasks, it can be expected that the tempo component of the output is of less relevance, for example when investigating metrical properties. Therefore, max-pooling can be applied across frequency in step F. In the example in the figure, the max-pooling was applied across the complete frequency vector.

However, tempo is a quasi-invariant musical property – the perceived tempo of a rhythmical pattern only varies somewhat linearly (in a certain range) with the speed at which that rhythmical pattern is performed. If the drum pattern presented in pane B of Figure 1 is performed very slowly, the pulse of the music may be perceived at the level of the hi-hat. But if the same pattern is performed much faster, say at a four times higher tempo, the perceived pulse level may instead be aligned with the kick and snare, or only with the kick drum. An identical activation pattern in D, differing by several octaves, will, therefore, have a slightly different perceptual and musical meaning. Because of this, it can be expected that better results are achieved for many tasks with a max-pooling operator that does not span the entire spectrum. Dividing the pooling operator into bands of one octave, for instance, gives a more expressional network. When processing the log-frequency representation in many layers, the pooling can be performed at smaller frequency ranges for each layer, so that step E and F are repeated several times (as indicated by the arrow in Figure 1). Pooling may also be useful to compensate for the fact that gradual tempo changes may smear the response of the CQT to different extents at low and high frequencies (more in low frequencies).

*Across time*

Average pooling across time (between time frames) and frequency can also be useful at this stage as a smoothing operator to reduce the variance in temporal fluctuation. For many tasks, such as tempo estimation, it may be suitable to compute the average across time for all activations if the tempo is stable. This could happen at step F, which would allow the system to account for variation in rhythm patterns across time (e.g., refrains and verses), before taking the average of the computed activations. It can also happen directly in conjunction with step C. This would make subsequent processing faster, but then the periodicities of different channels between different parts of the music will be smeared.

**G**

The output of the max-pooling across the whole spectrum is shown in pane G of Figure 1, for the three channels and four filters. As shown, the invariant rhythm processing has resulted in a very similar representation of music rhythm for the two tempi. If the random noise added to the two tempi had been identical, the output would also have been identical. However, the output representation can be expected to vary between the same rhythm patterns performed at different tempi. Music performers generally adapt accentuations to tempo, but the attack/decay of drums, for instance, will generally not vary according to tempo[1].

**H**

The rhythmical representation can be further processed before a final output is predicted, as shown in step H. A fully connected layer may be useful at this stage for high-level inference on the complete feature map. Step H may be applied for predicting overall aspects of rhythm, such as style, genre, groove, rhythmical complexity or the meter of the music.

## 2.2. Potential tasks

There are several types of rhythm tasks that can be solved by the methodology:
- The output of the log-frequency filter convolution can be used directly (after the activation function between step E and F) for rhythm tasks where the goal is to estimate the length of repetitions in the music. The beat length can be estimated to produce the tempo, the measure length can be estimated as it is an important metrical property, and the cepstroid can be estimated when the most salient periodicity is sought. To give an example, a tempo annotation can be converted to a beat length, and the correct beat length annotated as true while other beat lengths are annotated as false. A filter can be used in the last convolutional layer that is connected to all neurons of the previous layer for the same frequency. This filter is convolved across frequency. During runtime, the frequency with the highest output activation is used for the tempo prediction. As discussed in Section 6, a smooth filter can be learned in a final layer to weight predictions according to a desirable tempo distribution.
- After max-pooling, features such as those shown in pane G of Figure 1 can be used to compute more general concepts related to the rhythm of a track.
    o These can be used to estimate, e.g., *meter* and *swing*, rhythm related *genre* properties (i.e., classify different dance music patterns), and more general perceptual features such as "*groove*." A related system is the method proposed by Quinton (2017) for computing metrical modulations using NMF, coupled with a hidden Markov model.
    o Another potential perceptual feature is *Rhythmic complexity*, studied by Friberg et al. (2014). With the methodology suggested in Figure 1, the filter kernels can pick up energy distributions that deviate from the octave level and other important rhythm harmonics. These deviations will imply an increase in rhythmic complexity.
- Pattern matching for rhythms is a task in MIR that has been explored for both audio data (Klapuri, 2010) and symbolic music data (Janssen et al., 2013). By applying the log-frequency transform to shorter rhythm patterns, similar patterns performed at different tempi can be matched. These patterns will be manifested as frequency-shifted copies in D and can be represented by similar values in G after filtering in E and max-pooling in F. Therefore, they can be clustered. As the representation in D is invariant with regard to rhythmical phase, the starting point for the excerpt will have less importance as well. This has previously been pointed out for linear-frequency representations of rhythm (Klapuri, 2010).

---

[1] However, mixing engineers generally do change the reverb length of the snare drum according to tempo.

Processing for tasks proposed above can be enhanced by using relative phase as described in the next section. Note that the information given in the input of the specific example does not activate at the length of downbeat positions (assuming 4/4 meter), which is also reflected in pane D. The measure level may, however, be predicted through the log-frequency convolutional layers, which could infer that this rhythm represents a 4/4 meter with a measure length at half the lowest activated periodicity frequency.

## 3. Inference via magnitude and relative phase

### 3.1. On the use of phase information

Omitted from pane D in Figure 1 is the phase information ($\varphi$) of the transformation. When the input channels code for individual rhythmical elements such as the kick and snare drum, the phase will describe how these elements co-occur. Therefore, phase information is very important and should improve performance for most tasks. The *"unwrapped" absolute difference in phase* (the "*phase alignment*") between two channels can be used as a measure of co-occurrence of various rhythmical elements. This measure is useful, as it is invariant with regard to the phase of the rhythm at the investigated time-point (for example, the phase alignment will be the same at the second and third beat). The "raw" phase output from the CQT instead provides a direct mapping between frequency magnitudes and time domain magnitudes, expressed through windowed sinusoidal waves in the inverse transform. A computed activation pattern in the log-frequency domain can therefore be mapped to various magnitudes at specific time points if the raw phase is retained and used at a later stage. The raw phase has previously been used for returning to the time domain after filtering in harmonic/percussive source separation (FitzGerald, 2010; Elowsson and Friberg, 2015), but also for beat tracking (Elowsson, 2016). In the context of the methodology presented in Section 2, such mappings are necessary for tasks that deal with time-related predictions, such as beat and downbeat detection. This subsection deals with both relative and raw phase representations, and how they can be combined.

## 3.2. Relative phase

The unwrapped absolute phase difference, the *phase alignment* $x_{AB}$, between two channels $A$ and $B$, for a specific log-frequency $f$, can be computed according to Eq. 1,

$$C = |\varphi_{Af} - \varphi_{Bf}|, \qquad x_{AB} = \begin{cases} C, & \text{if } C < \pi \\ 2\pi - C, & \text{if } C \geq \pi \end{cases}, \qquad (1)$$

where $\varphi_{Af}$ and $\varphi_{Bf}$ corresponds to the phase of channel $A$ and $B$ at a particular frequency. The unwrapping of $C$ is used due to the circular properties of phase, to confine the phase alignment of two channels to the range $0$-$\pi$. The log-frequency magnitudes and phase differences can be sliced together in every other row of a convolutional layer. This is shown, for the same music rhythm processed in the left D-pane of Figure 1, in Figure 3. Magnitude and phase values have been scaled to cover the same range. The slicing of phase and magnitude allows for filters of height 4 (or, e.g., 2 or 6) to convolve the channel-frequency space, with a stride of two across the channel-dimension, as indicated by the green and cyan rectangles in Figure 3. The figure therefore assumes that kick, snare, and hi-hat encompass some sort of locality, i.e., that it can be relevant to analyze the relationship between the kick and snare drum by weights similar to those for the relationship between the snare drum and hi-hat. This was discussed in Section 2.1.E – "across channels." In cases where this does not hold, but phase relationships still are relevant, the filter would instead extend across all channels (including phase alignment-channels). Slicing of phase alignment-values is not necessary in that case.

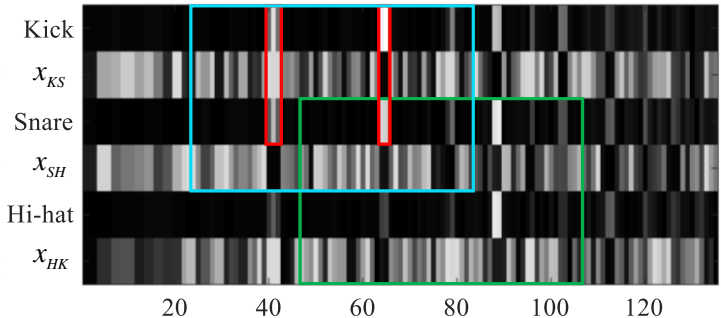

**Figure 3.** A representation filtered by convolutional filter kernels in the log-frequency domain; the representation consisting of magnitudes and the unwrapped phase difference between neighboring channels (the phase alignment). Magnitude values were rescaled for illustrative purposes. Red rectangles highlight antiphase at the half-measure level and phase alignment at the beat level between kick and snare ($x_{KS}$). Sizes of convolutional filters are illustrated by cyan and green rectangles.

The stride across frequency should normally be one, because the resolution is better regulated by adjusting the number of bins per octave in the frequency transform (this is true for the methodology described in Section 2 as well). The variation in phase alignment between the kick and snare drum at different frequencies is indicated by red rectangles in the figure. The kick and snare have a maximum phase separation ($\pi$) at twice the beat level, but are phase-aligned (0) at the beat level.

Phase alignment values are only relevant when the corresponding magnitudes of both channels are reasonably high. In other cases, these values are mostly noise. The network will use non-linear activations and can therefore account for this in the processing. But to be able to do this efficiently together with other tasks, it can be expected that more than one convolutional layer is needed in the log-frequency domain (steps D-E-F in Figure 1).

Eq. 1 is useful when comparing phases belonging to channel bins with the same frequency. As touched upon earlier, rhythmic elements commonly interact at integer multiples. Subdivisions often have a 2, 3, or 4 times higher frequency than the beat (i.e., there are 2, 3, or 4 subdivisions per beat), and beats often have a 2, 3, or 4 times higher frequency than the measure level. The relevant periodicities (subdivisions, beats and slower periodicities) are, therefore, to a large extent related as integer multiples. Given a unit frequency $f_1$, important rhythmically related higher frequencies can be found at integer multiples

$$f_m = \{2, 3, 4, 6, 8\}. \qquad (2)$$

The phase alignment $x$ between $f_1$ with frequency $\omega_1$ and phase $\varphi_1$, and any integer of $f_m$ with frequency $\omega_m$ and phase $\varphi_m$ can be computed as

$$C = \left|\varphi_1 - \varphi_m \frac{\omega_m}{\omega_1}\right|, \qquad x = \begin{cases} C, & C < \pi \\ 2\pi - C, & C \geq \pi \end{cases}. \qquad (3)$$

Added from Eq. 1 is the multiplication by the corresponding frequency ratio of the phase bins. The equation can be understood in the following way: move the two phasors along the unit circle according to their relative frequencies until the slower phasor has $\varphi = 0$. At that point, the unwrapped absolute phase of the faster phasor corresponds to the phase alignment.

An illustration is presented in Figure 4. The figure visualizes how the log-frequency transform responds to sounds in two different channels, at relevant frequencies. The phase is measured at time-point Y, and Eq. 3 is applied. The computed phase alignment corresponds to the angle between the two phasors at time-point Z, when the phase of the lower frequency phasor is 0.

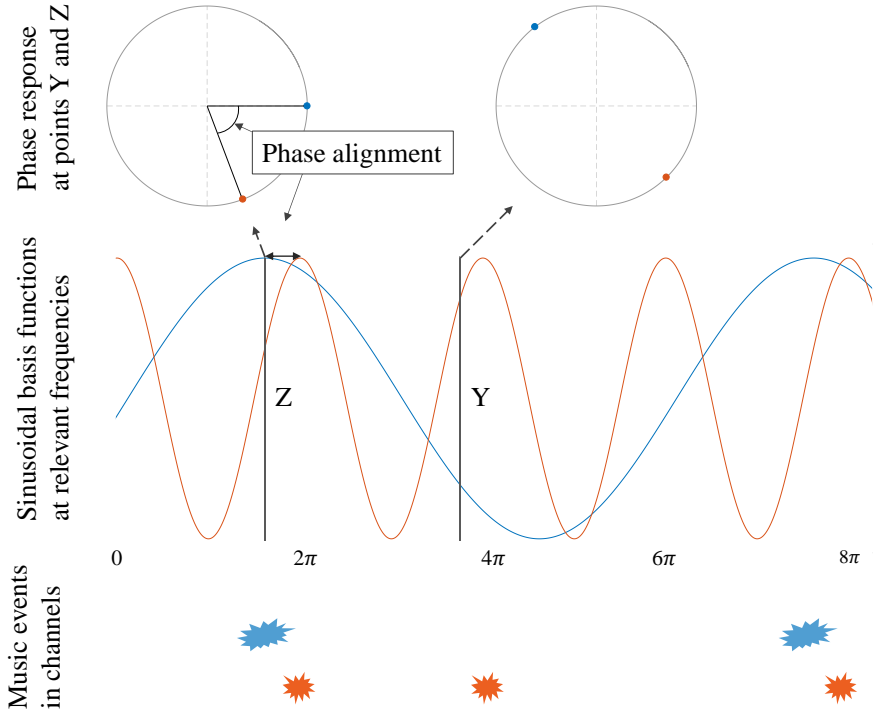

**Figure 4.** Channel activations (bottom) generate frequency responses from the CQT that can be visualized with sinusoidal curves across time (middle). The phase (top) for the two frequencies (blue and red) is measured (time-point Y). The phase alignment is computed, which corresponds to the angle between the higher-frequency phasor (red) and the lower-frequency phasor (blue) when it is at 0 (time-point Z).

The figure illustrates two frequencies that are not very well phase- aligned. This implies that the two frequencies (periodicities) do not both have an important function (e.g., subdivisions, beat level, measure level) in the metrical structure – periodicities that are perceptually important are generally phase-aligned. High-magnitude frequencies may be unaligned for various reasons. The log-frequency transform may have reacted to spurious activations, for example from vocal fricatives, or syncopated musical elements, such as a kick drum or an expressive vocal performance. Given the lower phase alignment, the network can be expected to assign a lower probability during processing that both frequencies are musically important.

Eq. 3 can only be used to determine a unique phase alignment when $f_m$ is an integer multiple of $f_1$, i.e., when the division can be reduced to a unit fraction. For other integer fractions $n/m$, such as 2/3, the phase alignment will vary over the $n$ periods that the lower frequency completes while the higher frequency completes $m$. Phase alignment can then be measured by computing the angle when the phase of the lower frequency is 0 for all $n$ periods. The minimum phase alignment over the $n$ periods could then be used, as long as $n$ is a small integer. This would be complicated to integrate into gradient descent, and hardly worth the development effort. Therefore, it should only be useful in the second case, when the deep learning starts at step D in Figure 1. As *n* gets larger, the phase alignment value will, however, have less and less importance.

Phase alignment values from Eq. 3 for $f_m$ can be inserted with magnitudes by slicing them in, just as for the basic phase alignment measure sliced with magnitudes in Figure 3. If using $f_m$ (five additional phase alignment-values), each channel bin will be increased in size from 2 to 7 in the CNN processing. The stride across channels will need to be increased accordingly.

### 3.3. Improving generalization

Some frequency bins at a distance from each other that does not correspond to any musically meaningful level of repetition may be of negligible importance. In this case, a sparser structure can be used, similar to how only harmonics of fundamental frequencies are processed in recent polyphonic transcription implementations (Bittner et al., 2017; Elowsson, 2018a). Of course, log-frequency representations of rhythm also have these harmonics at fixed distances (with a somewhat different distribution), that emerge as a result of the integer relationship of various repetitions in music. A CNN for processing only "rhythm harmonics" can be created by stacking frequency-shifted copies of the log-frequency representation (i.e., a fourth dimension), then using filters that extend across this stack but do not extend across frequency. This strategy has been used for fundamental frequency estimation (Bittner, 2017). The frequency shift *s* is applied to all specified harmonics, e.g., *h* = {1 2 3 4 6 8}, as

$$s = \lfloor \log_2(h) \times bpo \rfloor. \tag{4}$$

The operation $\lfloor x \rfloor$ is used to round the matrix shift to the closest integer and *bpo* represents the number of frequency bins per octave. Appropriate frequency shifts can be improved by using techniques similar to those proposed by Elowsson (2018a) – learning which frequency positions are the most useful for performance. The range could also be expanded, e.g., using *h* = {9 12 16}. The stacked frequency shifted matrices will use kernels that span across the depth of the stack, potentially a few channels, with a stride of one across frequency and a stride across channels adapted to the number of phase alignment bins used (as previously specified).

### 3.4. Transforming back to the time domain using stored phase

The raw phase from the CQT (step C in Figure 1) can be saved, and then used after the convolutional layer(s) to convert log-frequency activations to time domain activations via the inverse CQT (ICQT). It is rather likely that the raw phase of one or several channels from the CQT transformation will provide a useful mapping to the time domain, without processing the phase further. Such a processing scheme is used by Elowsson (2016). In that study, the CQT was computed for the beat activation signal. A filter was applied to the magnitudes, setting all magnitudes outside of a selected "tempo-frequency" to 0, while leaving the phase untouched. When transforming back to the time domain using the processed magnitude and unprocessed phase, a smooth sinusoidal wave was produced containing peaks aligned with the annotated beat positions.

The same methodology can be used, for example, for beat tracking and downbeat tracking within the proposed system. For beat tracking, the system would learn to activate at the frequency corresponding to the tempo before inverting to the time domain, and for downbeat tracking, the system would learn to activate at the frequency corresponding to the measure length. Targets in the frequency domain can also be computed from time domain annotations as discussed in Section 3.5.

The methodology can be used for rhythm tracking in the proposed system under the assumption that the phase information is relevant. Relevance is context dependent; for the beat tracking task, it would imply that bin-phases map magnitudes to the time domain so that resulting peaks are aligned with the beat positions. The extent to which this mapping is correct depends on the input to the CQT step. If the learning of the network starts after the CQT (step C in Figure 1), appropriate pre-processing must be determined through experimentation across varying audio examples. If the learning of the network starts before the CQT, the network will need to learn how to process the low-level time-frequency representation of the audio, to produce activations for relevant rhythmical features with a useful phase.

There is another aspect to also consider when transforming back to the time domain. Should a full concatenation of channels occur within the CNN (e.g., with a final fully connected layer between channels) so that only a single channel remains to be inverted, or should several channels be invertible? Or should perhaps several of the previously computed raw phases be applied to a single channel of activations, and then weighted together after inversion? The channel(s) with the most relevant phase response can be determined in the log-frequency domain, based on extracted activation magnitudes and phase alignments. This variation provides the network with options. If several channels are inverted, the convolutional layer(s) can determine the channel(s) with the most accurate phase in the frequency domain and activate strongly on those channels. For example, for a song with a syncopated kick drum pattern and a steady piano accompaniment, the network can determine that any channel activating for the pitched sounds, at the pitch range and panning position of the piano, provides a reliable phase response.

## 3.5. Training with time domain targets

The approach suggested for association to the time domain from raw phase values in Section 3.4 can be extended or complemented by using time domain targets during training.

The CQT with the same parameters as in step C of Figure 1 can be applied to the time domain annotations (e.g., beat annotations) and the resulting frequency-domain response in each time frame can be used as a ground truth annotation for training. The benefit of this method is that it offers a natural way of letting the annotations smoothly adapt to tempo variations when represented in the frequency domain. Therefore, the method is useful both for beat tracking and downbeat tracking as well as tasks where estimates are computed directly in the frequency domain, such as tempo estimation.

The simple solution is to just insert a one at the time points of annotations in a vector that is otherwise all zeros and compute the short-time CQT across that vector. This would, however, create disturbances and harmonics in the frequency response. A better solution is therefore to generate a sinusoidal curve with peaks at beat annotations, before computing the CQT of this curve. This is therefore suggested as the main method for creating training targets whenever time domain annotations such as beats are available for relevant tasks. After the CQT is applied, the targets will consist of a vector *TA* of magnitudes but also phases for each time frame, the same size as the frequency vectors produced in step C and shown in pane D of Figure 1. The magnitudes can be normalized to a desirable range for training, e.g., 0-1. Only the phase at true target frequencies is relevant.

If previous phase values are used for inverting to the time domain for e.g. beat tracking, as outlined in Section 3.4, the phase may be:
- discarded
- or, whenever the phase responses of several channels are applied as described in Section 3.4., it can be used to annotate the accuracy of the phase responses of the channels at the correct frequency bin (the bin the corresponds to the annotated frequency). By this strategy, it is still possible to only use relative phase during prediction.

In Section 4, a method will be suggested that can use the "raw" phase values during predictions: employing complex-valued CNNs. The *TA* can then be used directly as an annotation, training the network to predict magnitudes and phases of the *TA*. For the phase, the system can restrict the scope, and only train for phases at bin frequencies where the *TA* is annotated as true or is above a certain threshold, tuned in a parameter sweep.

# 4. Inference using complex-valued networks

## 4.1. Complex-valued neural netwroks

Several restrictions were enforced on the network by using real numbers (phase alignment in combination with magnitude), to keep the architecture simple and promote high generalization. A more expressive way of dealing with wave phenomena is complex-valued neural networks (Hirose, 2012). The relationship between rhythmic elements can be described by phase and amplitude in the complex plane. When using the complex output of the CQT, and processing the data with complex-valued neurons, the complex-valued relationship between channels can be preserved during processing. At the same time, the input values are not relative, i.e., the input does not simply describe relative phase alignments between channels, but the actual phase of each bin. The frequency of each bin is represented by its bin-frequency position.

These properties in combination mean that the network should be able to compute, for example, a beat activation, without applying the ICQT with original phase values. In other words, it would no longer be necessary to let the previously computed phase imply how activations for a specific frequency should be distributed across time. It may, however, still be desirable to do so, by instead trying to predict the *TA* outlined in Section 3.5 for each time frame and inverting to the time domain with the *computed* phase. The benefit of this method in relation to computing a simple beat activation would be that predictions will affect the beat activation across a wider time window, producing a smooth sinusoidal beat activation curve reducing risks of noisy predictions.

Networks with complex-valued neurons have been used for a long time, dating back to at least 1998 for MIR (Kataoka, Kinouchi, & Hagiwara, 1998). If locality is important, the complex-valued CNN, CV-CNN, offers a way of dealing with both phase and locality. A CV-CNN has previously been used for polarimetric SAR image classification (Zhang et al., 2017), where the input data both have locality and phase.

Figure 5 illustrates the relationship between the two rhythmical elements, A and B, and a mapping between these elements described by real numbers (red line) or through phase and amplitude (green line). The rhythmical pattern B has a higher amplitude than A, and the periodicity of the B-pattern has a phase offset of roughly $-\pi/2$. The rhythm patterns are illustrated at two different time-points (Map1 and Map2), showing that the complex valued mapping is invariant with regard to raw phase.

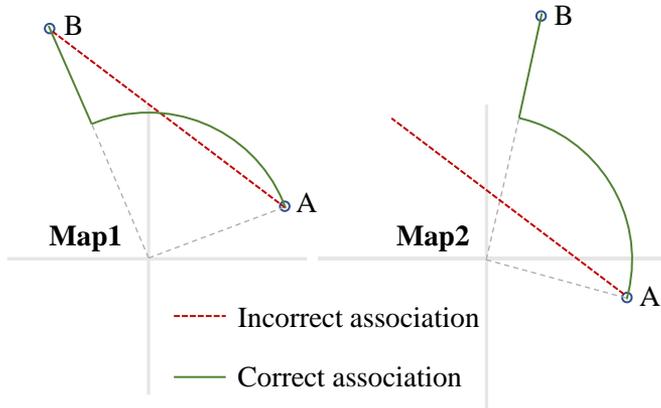

**Figure 5.** Two rhythmical patterns associated via phase and amplitude or two real numbers. The complex-valued mapping is invariant with regard to phase.

In essence, when using complex-valued processing, the relationship between rhythmical elements can be preserved across time, whereas processing by real numbers will fail to find a mapping that generalizes to phase shifts. This has been illustrated in the more general case of complex-valued NNs by Hirose (2012) in fig. 3.2-3.3 (similar to Figure 5).

## 4.2. Frequency

For bins of different frequencies, phase plus amplitude is not enough information to represent phase alignment. However, if the frequency of the two bins, or more specifically the log-frequency interval of the bins, is also included, enough information is provided. The frequency interval is implicitly encoded by distance in the kernel across frequency during convolution. Phase alignment involving frequency was explored in Figure 4. There is, however, another aspect of phase where the log-frequency interval between bins is not enough. It could be very useful to use the phase (not relative phase) in combination with the frequency to determine the time offset from the next implied time domain peak (phase 0). The time domain peak in Figure 4 for the lower frequency sinusoid is marked with Z. The time distance to this peak from point Y in the figure (distance to the first previous peak) can be computed from

$$\omega\varphi. \qquad (5)$$

Moreover, the distance to the next peak can be computed from

$$\omega(2\pi - \varphi). \qquad (6)$$

Eqs. 5-6 assumes that the log-frequency transformation was "centered," as previously mentioned. Evidently, complex-valued neurons have relevant properties for rhythmic processing. There are frequency-related properties of music that provide reasons to explore how frequency can be included in complex-valued networks. By encoding bin values as a triplet, {amplitude, phase, frequency}, the time domain association is made clear. How this can be incorporated in complex-valued networks, and further exploration, is however left for future work.

## 5. Task overview

This article has presented various methods for rhythm processing in a generalized tempo-invariant fashion. Some of the tasks solved without max-pooling across the tempo range are shown in Table 2. The process for each task is described by the corresponding entries in the headers of each row and column; i.e., downbeat tracking is performed by producing activations at the measure length and inverting to the time domain using the stored phase.

| *Produce activations at:* | *Beat length* | *Measure length* |
|---|---|---|
| Compute the maximum frequency activation | **Tempo estimation** | **Measure length Estimation** |
| Invert to time domain with stored phase | **Beat tracking** | **Downbeat Tracking** |

**Table 2.** The last processing step and the associated periodicity-frequency for four important rhythm tracking tasks.

Some of the tasks solved after max-pooling across frequency in step F of Figure 1 are shown in Table 3. In the first column, the main functionality of the tempo-invariant processing step is described for a task named in the second column.

| *Tempo-invariant main functionality performed in convolutional layers(s)* | *Task/Estimating* |
|---|---|
| Filtering to detect style-specific patterns | **Music genre/ Dance styles** |
| Filters extract distinguishing rhythmical features by which patterns are identified | **Pattern Matching** |
| Analyze the spread and phase alignment of activations across frequency | **Rhythmic Complexity** |
| Analyze activation strength for channels that cover various aspects of the audio | **Rhythmic clarity** |
| Analyze relationship between periodicities in different channels | **Meter** |
| Extract distinguishing features; audio at different speeds will be represented by the same values after max-pooling | **Copyright infringement** |
| Detect phase and magnitude variations produced by swing variations | **Swing** |

**Table 3.** Tempo-invariant functionality used for a wide variety of tasks that can be solved after max-pooling across frequency.

# 6. Additional remarks

The CQT implementation used for showing the example in Figure 1 has an optional input parameter for controlling the time-frequency trade-off in lower frequencies. A better time resolution can be achieved at lower frequencies by setting this parameter to a non-zero value. This could be useful for songs with tempo drifts to avoid a smudged response in lower frequencies.

As touched upon in Section 2.1.F, tempo is a quasi-invariant property. Forcing the network to process rhythm patterns with the exact same weights several tempo-octaves apart will reduce performance. Therefore, an intermediate layer could be applied where each neuron is multiplied by a weight, to tune the response across frequency. Weights that diverge from neighboring weights at lower and higher frequencies for the same weight vector are penalized during learning, to enforce a smooth response across frequency. Similar strategies can be useful for processing other quasi-invariant aspects, such as pitch. The author has previously tried such an approach for HPSS with good results (no publication available).

Another strategy for handling the quasi-invariant property of tempo is to compute the maximum value *and* its position during max-pooling across the complete tempo range. The output of each such pooling would then consist of a tempo-invariant filter response as well as the tempo at which that response occurred. This is a rather expressive yet compact way of describing rhythm, something that should promote generalization. The position-value will however not change smoothly, which may create problems in many training scenarios. As an alternative, the position-value can therefore be added after initial training is completed, and training redone separately from step H outlined in Section 2 and Figure 1.

# 7. Acknowledgements

Thanks to Andreas Selamtzis for valuable assistance with Eq. 3. Thanks to Anders Friberg, Pawel Herman and Tony Lindeberg for manuscript comments.